\documentclass{article}
\usepackage{stywhispers,amsmath,epsfig}
\usepackage{graphicx}
\usepackage{amssymb}

\newcommand{\N}{\mathcal{N}}

\title{HS-ANET: Star Spectral Type Enhanced Astrometric Calibration for Hyper Spectral Space Imaging} 
\name{William Mitchell$^{*,1}$\thanks{* Co-authorship}, Kevin Phan$^{*,2}$, David Chaparro$^{2}$, Enrique De Alba$^2$, J. Zachary Gazak$^3$}
\address{$^1$The University North Texas, 1155 Union Cir, Denton, TX 76205, USA \\
$^2$EO Solutions, 6713 South Eastern Ave., Las Vegas, NV 89119, USA \\
$^3$U.S. Space Force, Space Systems Command, 550 L\={\i}poa Parkway, Kihei, HI 96708, USA}
\begin{document}
%

\maketitle
\begin{figure*}[!ht]
  \includegraphics[width=\linewidth]{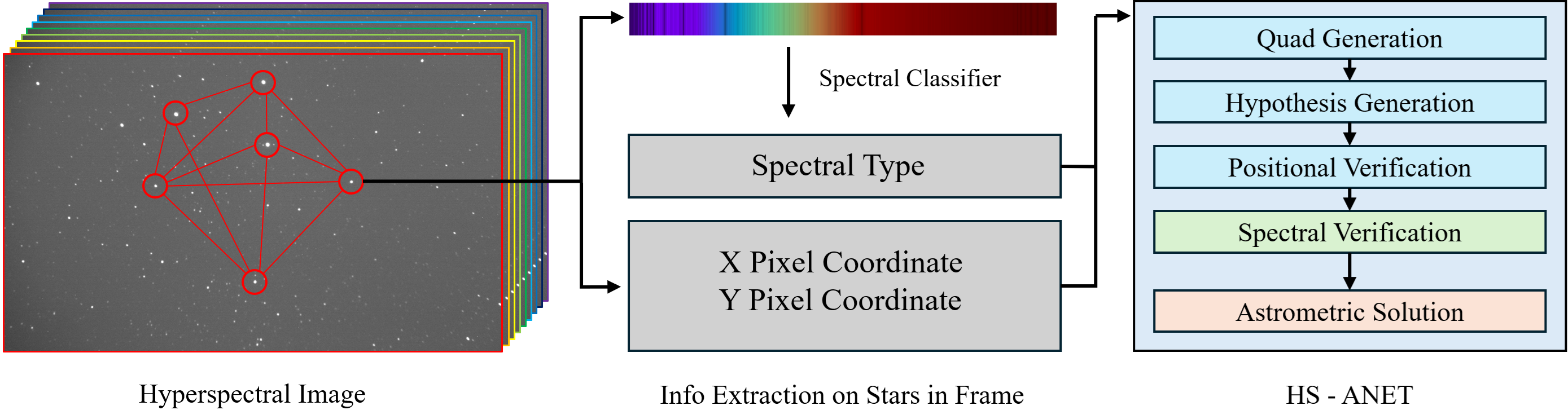}
  \caption{Proposed pipeline in which HS-ANET can be used to optimize low field of view telescope astrometric fitting.}
\label{fig:pipeline}
\end{figure*}
\begin{abstract}
Traditional lost-in-space algorithms, such as those implemented in astrometry.net, solve for spacecraft orientation by matching observed star fields to celestial catalogs using geometric asterisms alone. In this work, we propose a novel extension to astrometry.net that incorporates stellar spectral type, which is derived from hyperspectral imagery, into the matching process. By adding this spectral dimension to each star detection, we constrain the search space and improve match specificity, enabling successful astrometric solutions with significantly fewer stars. Our modified pipeline demonstrates improved fit rates and reduced failure cases in cluttered or ambiguous star fields, which is especially critical for autonomous space situational awareness and traffic management. Our results suggest that modest spectral resolution, when incorporated into existing geometric frameworks, can dramatically improve robustness and efficiency in onboard star identification systems.
\end{abstract}

\section{Introduction}
\label{sec:intro}
The escalating congestion of Earth's orbital environment poses significant challenges for space operations. As of 2024, over 35,000 objects larger than 10 cm are tracked in Earth's orbit, with estimates suggesting over 1 million debris pieces between 1 cm and 10 cm in size. This proliferation of space debris increases the risk of collisions, threatening both active satellites and future missions. The situation is further exacerbated by the surge in satellite deployments, with 2,578 operational satellites launched in 2024 alone \cite{ESA2024}.

Accurate localization of objects in space is crucial for collision avoidance and effective space traffic management. Astrometric solutions, which determine the position and orientation of spacecraft or debris by matching observed star fields to known catalogs, are fundamental to this process. However, traditional astrometric methods face limitations, especially when using narrow-field-of-view (FOV) telescopes \cite{Felt2023, fritz2010limiting}. These instruments, while offering higher resolution and increased precision in Right Ascension (RA) and Declination (DEC), often capture fewer stars in a single frame, complicating the matching process. 

To address this challenge, we propose an enhancement to the widely-used astrometry.net (ANET) algorithm \cite{lang2010} by incorporating stellar spectral type information derived from hyperspectral imaging. Hyperspectral imaging telescopes capture detailed spectral information across numerous narrow, contiguous wavelength bands, allowing for the precise analysis of stellar spectra. By examining the absorption and emission lines within these spectra, astronomers can determine the chemical composition, temperature, and other physical properties of stars, leading to accurate classification into spectral types such as O, B, A, F, G, K, and M \cite{Mantelet2016}. 

This classification is essential for understanding stellar evolution and for improving astrometric solutions, especially in scenarios where traditional imaging may not provide sufficient data due to a limited number of visible stars. Hyperspectral sensors, which can act as an alternative to traditional spectrographs, offer the advantage of capturing spatial and spectral data simultaneously, enabling the detection of key spectral features like hydrogen Balmer lines, TiO bands, and metallic absorption lines across an entire field of view \cite{Specim2024}. 

We propose Hyperspectral ANET (HS-ANET), in which spectral data is utilized in astrometric algorithms to enhance the ability to match observed stars with cataloged counterparts.  Enabled by hyperspectal imaging, Figure \ref{fig:pipeline} depicts the use case of HS-ANET as a downstream algorithm for enhanced astrometry. By leveraging the rich spectral information provided by hyperspectral imaging, we reduce the number of stars required for an astrometric fit, facilitating accurate spacecraft localization and contributing to effective space traffic management.

\section{Data Simulation and Sourcing} 
\label{sec:format}

\begin{table}[ht]
\footnotesize
\centering
\begin{tabular}{llll}
\hline
\textbf{Feature} & \textbf{Tycho-2} & \textbf{Gaia DR2} & \textbf{Gaia DR3} \\
\hline
Release Year & 2000 & 2018 & 2022 \\
Number of Stars & $\sim$2.5 million & $\sim$1.7 billion & $\sim$1.8 billion \\
Astrometry (RA, Dec) & $\sim$60 mas & $\sim$25 $\mu$as & $\sim$25 $\mu$as  \\
Spectral Types & No & Limited  & Yes  \\
Spectra Availability & No & No & Yes  \\
\hline
\end{tabular}
\caption{Comparison of Tycho-2, Gaia DR2, and Gaia DR3 star catalogs}
\label{tab:catalog_comparison}
\end{table}

\subsection{Star Catalogs}

Modern blind astrometry tools, such as Astrometry.net, perform geometric matching of detected star configurations (quads) against reference star catalogs. The standard implementation typically uses the Gaia DR2 catalog \cite{GaiaCollaboration2018}, supplemented with Tycho-2 \cite{Hog2000} for improved coverage of bright stars. This pairing serves as the baseline in our evaluation and reflects common practice in high-accuracy astrometric systems.

The Tycho-2 catalog contains approximately 2.5 million of the brightest stars ($V \lesssim 11$), with astrometric and photometric measurements derived from the Hipparcos mission. Despite its age and lower precision, Tycho-2 remains useful in legacy systems due to its completeness for bright stars. Gaia DR2, released in 2018, significantly expanded sky coverage with over 1.7 billion stars and provided milliarcsecond- to microarcsecond-level precision in astrometry and proper motion. However, it lacked spectral classification data and low-resolution spectra.

In contrast, our method utilizes Gaia DR3, which has numerous advantages outlined in Table \ref{tab:catalog_comparison}. This includes enhanced spectrophotometric measurements via the BP (blue photometer) and RP (red photometer) bands, as well as derived astrophysical parameters such as stellar effective temperature and spectral type classification. Gaia DR3 effectively subsumes Tycho-2 in coverage, containing approximately 98.4\% of all stars in the Tycho-2 catalog, and therefore does not require explicit supplementation \cite{GaiaDR3}.

\subsection{SatSim Data}

\begin{figure}[bh!]
    \centering
    \includegraphics[width=1\linewidth]{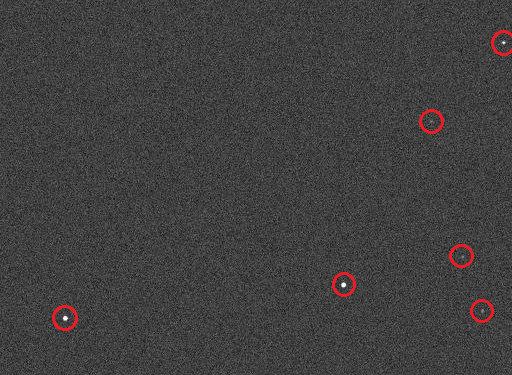}
    \caption{Example of real world accurate starfield generated by SatSim with FOV set to 0.25 degrees. Stars are circled in red and come with a spectral type annotation which can be used for performance benchmarks for HS-ANET.}
    \label{fig:enter-label}
\end{figure}

To generate realistic astronomical scenes for testing and evaluation, we utilize a modified version of SatSim, a starfield and space scene simulator \cite{cabello2022satsim}. SatSim is designed to support the development of space situational awareness (SSA) and astrometric algorithms by rendering high-fidelity images of the night sky based on satellite-relative viewpoints. It models camera characteristics such as field of view, resolution, and point spread function (PSF), and can simulate a wide range of orbital conditions. SatSim is ideal for validating algorithms in scenarios with limited star visibility, varying noise levels, or motion blur.

To enable spectral-type-aware astrometric experimentation, we modify SatSim to use Gaia DR3 as the underlying star catalog. We simulate scenes where each star is annotated with both position and spectral classification to evaluate the impact of spectral verification under controlled conditions. SatSim provides reproducible testbed for assessing improvements in astrometric fit quality and solution robustness.


\section{Methodology}
\label{sec:methodology}

Astrometry.net performs four steps to find an astrometric fit: source extraction, quad generation, hypotheses generation, and positional verification. We propose a final spectral verification step to enhance the pipeline.

\subsection{Source Extraction}

Before astrometric solving can begin, the input image is processed to detect bright point sources, typically stars. These sources are extracted using centroiding and thresholding algorithms, resulting in a list of 2D coordinates $(x_i, y_i)$ corresponding to detected sources in the image. In our experiments, we assume that stars have been extracted with their spectral type for the preceding steps.

\subsection{Quad Generation}

From the extracted sources, the system constructs asterisms, specifically quads, sets of four stars selected to form a geometric configuration that is invariant to scale, rotation, and translation. These quads are described by a set of invariant parameters that allow for efficient indexing and comparison against a precomputed database of catalog quads.

\subsection{Hypotheses Generation}
Each quad found in the image is used to query a hashed index of catalog quads, producing a list of candidate matches. Each match corresponds to a pose hypothesis, a proposed world-to-image transformation that maps catalog celestial coordinates to image coordinates. 

\subsection{Positional Verification (Bayesian Decision Process)}

Once a candidate alignment hypothesis is generated, the system evaluates its plausibility using a Bayesian decision process. This step determines whether the alignment should be accepted or rejected by comparing the probability of the observations under two competing models:

\begin{itemize}
    \item \textbf{Foreground model} ($F$): assumes the hypothesis is correct—the observed stars in the image correspond to catalog stars under the proposed alignment.
    \item \textbf{Background model} ($B$): assumes the hypothesis is incorrect—the observed stars are unrelated to the catalog.
\end{itemize}

Let $D$ denote the set of observed detections in the image. The system computes the posterior odds of the alignment being correct as:

\[
\frac{P(F \mid D)}{P(B \mid D)} = \frac{P(D \mid F)}{P(D \mid B)} \cdot \frac{P(F)}{P(B)}
\]
\noindent
Where the first term is the Bayes Factor $K$. With the set of hypothesized alignment reference stars $\theta$ and individual star detections $t_i$, $K$ is defined as:
\small
\[
K = \frac{P(D \mid F)}{P(D \mid B)} = \sum_{i=0}^{|D|}\frac{P(t_i\mid F)}{P(t_i\mid B)} = \sum_{i=0}^{|D|}\frac{\sum_{j=0}^{|\theta|}P(t_i|\theta_j,F)P(\theta_j)}{P(t_i\mid B)}
\]
\normalsize

\noindent
The term $\frac{P(F)}{P(B)}$ is a manually set prior, which is set to $10^{-6}$ to be conservative \cite{lang2010}. The sequential foreground model is specified by its data likelihood and parameter prior: The likelihood of the data $t$ given parameters $\theta$ is: $P(t \mid \theta, F) = \prod_{i=1}^{|D|} P(t_i \mid \theta_i, F),$ where the probability for a single data point $t_i$ is given by:
$$P(t_i \mid \theta_i, F) =
\begin{cases}
1/A, & \text{if } t_i \text{ is a distractor} \\
\N(t_i \mid r_{\theta_i}, \sigma^2_i), & \text{otherwise}
\end{cases}$$
Where $A$ is the area of the image and $\N(x \mid \mu, \sigma^2)$ is the probability density of drawing $x$ from a Gaussian distribution with mean $\mu$ and variance $\sigma^2$.
The prior on the parameters $\theta$, $p(\theta \mid F)$, requires that each parameter assignment from a reference star to a detection is unique. We define it as follows:
$$ p(\theta \mid F) = \begin{cases}
0, & \text{if } \theta_i = \theta_j \text{ for any } i \neq j \\
\prod_{i=1}^{|D|} p(\theta_i \mid F), & \text{otherwise}
\end{cases} $$
The probability for an individual parameter assignment, $p(\theta_i \mid F)$, is a mixture model where $|D|$ is used to represent a uniform choice over the set of detections for a non-distractor:
$$ p(\theta_i \mid F) =
\begin{cases}
d + (1 - d) \cdot \dfrac{\mu_i}{|D|}, & \text{if } t_i \text{ is a distractor} \\
\dfrac{1 - d}{|D|}, & \text{otherwise}
\end{cases} $$

\normalsize
\noindent
Where $d$ is the assumed fraction of distractor stars. In a simple background model:

\[ P(t_i|B) = \frac{1}{A}\]

In order to accept a hypothesized $\theta$, the posterior odds must be larger than the expected utility of each decision outlined by the utility table defined by Astrometry.net and in Table \ref{tab:utility}. 

\[\frac{P(F\mid D)}{P (B\mid D)} > \frac{u(TN)-u(FP)}{u(TP)-u(FN)}\]

\begin{table}[ht]
\centering
\begin{tabular}{lcc}
\hline
\textbf{Decision} & \textbf{Hypothesis True} & \textbf{Hypothesis False} \\
\hline
Accept (Match)    & $+1$                     & $-1999$ \\
Reject (No Match) & $-1$                      & $+1$ \\
\hline
\end{tabular}
\caption{Utility table used in Astrometry.net's Bayesian decision process.}
\label{tab:utility}
\end{table}

From this, the required $K$ for a hypothesis acceptance can be calculated as follows:
\begin{align*}
K &> \frac{P(F)}{P(B)} \cdot \frac{u(TN)-u(FP)}{u(TP)-u(FN)}\\
K &> 10^6 \cdot \frac{1 - (- 1999)}{1 - (- 1)}\\
K &> 10^9
\end{align*}

\subsection{Spectral Verification}
Once a set of detections undergoes the positional verification process, we determine whether spectral verification is required and increment or decrement the Bayes Factor depending on spectral matching.

If a hypothesis yields a Bayes Factor $K$ that is promising but insufficient for acceptance ($\tau < K < 10^9$), we perform an additional spectral vertification step. The Bayes Factor is adjusted based on spectral consistency:
\begin{equation*}
    K \xleftarrow{} K + \sum_i^{|D \cap \theta|}
    \begin{cases}
+\lambda, & \text{if } \theta_i \text{ and } t_i \text{ are a spectral match} \\
-\lambda, & \text{otherwise}
\end{cases}
\end{equation*}
\\
\noindent
Here, $D \cap \theta$ is the set of all observed detections that positionally matched with a corresponding hypothesis star. The odds-to-tune (OTT) threshold ($\tau$) and the reward-penalty constant ($\lambda$) are tuned experimentally to filter false matches and achieve the most robust performance. The updated $K$ is then re-evaluated to determine if the hypothesis can be accepted.

\section{Results}
\begin{figure}[t!]
    \centering
    \includegraphics[width=\linewidth]{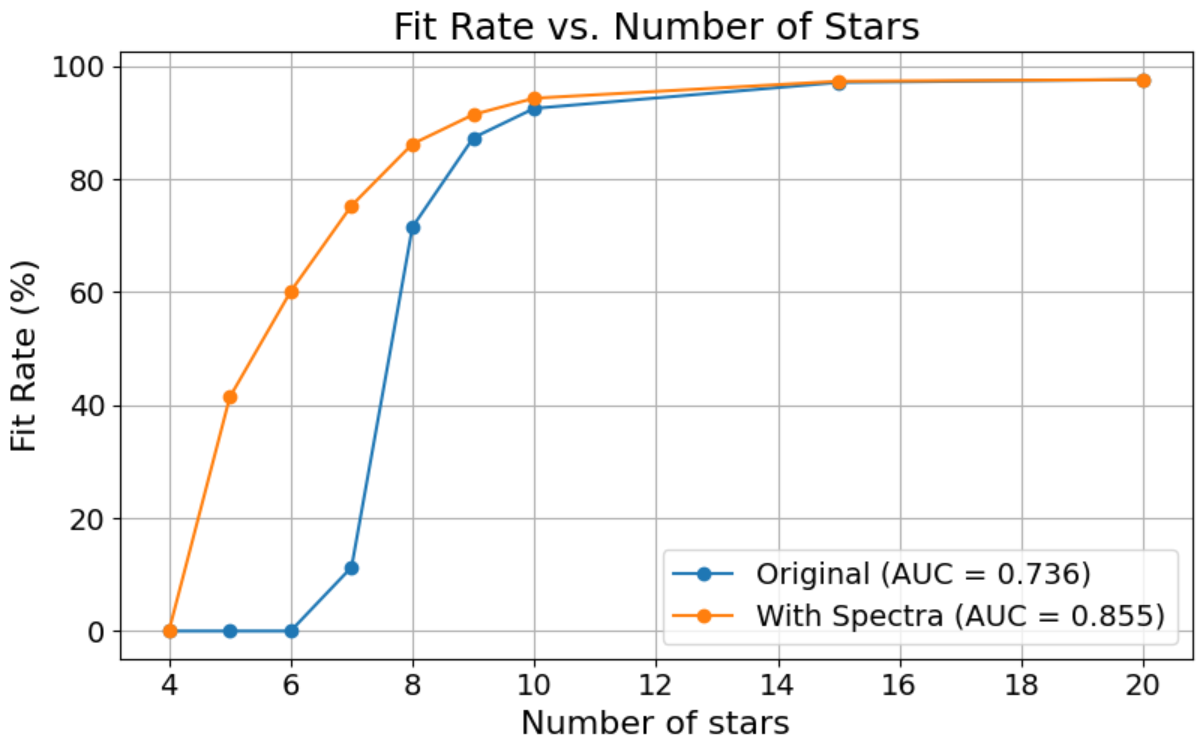}
    \caption{Fit rate vs. number of stars per image, comparing the original solver and HS-ANET.}
    \label{fig:line_graph}
\end{figure}

\begin{table}[ht]

\centering
\begin{tabular}{ccccc}
\hline
& \multicolumn{2}{c}{Fit Rate} & & \\
\cline{2-3}
\textbf{Stars} & \textbf{ANET (\%)} & \textbf{HS-ANET (\%)} & \textbf{$\lambda$} & \textbf{$\tau$}\\
\hline
20 & 97.6 & 97.6 & $10^2$ & $10^3$ \\
15 & 97.1 & 97.3 & $10^3$ &$10^3$ \\
10 & 92.5 & 94.3 & $10^3 $&$10^3$ \\
9  & 87.3 & 91.4 & $10^3$ &$10^3$ \\
8  & 71.5 & 86.2 & $10^2$ & ---\\
7  & 11.3 & 75.3 & $10^5$ &$10^3$ \\
6  & 0    & 60.1 & $10^4$ &$10^3$ \\
5  & 0    & 41.4 & $10^4$ &\\
4  & 0    & 0    & --- & ---\\
\hline
\end{tabular}
\caption{Fit rates for original and spectral methods by number of stars, along with the best method used.}
\label{tab:fit_rates}
\end{table}
\begin{table}[th!]
    \centering
    \begin{tabular}{cccc}
    \hline
    Star Range & ANET & HS-ANET & Performance Change\\
        \hline
       \textnormal{[10, 20]} & 0.974 & 0.974 & \textbf{+0.000}\\
       \textnormal{[4, 20]} & 0.736 & 0.855 & \textbf{+0.119}\\
       \textnormal{[4, 10]}  & 0.361 & 0.669 & \textbf{+0.308} \\
    \hline
    \end{tabular}
    \caption{AUC performance of ANET and HS-ANET across different star regimes.}
    \label{tab:auc}
\end{table}

To evaluate the performance of our hyperspectral-enhanced solver, HS-ANET, we generated a dataset of 958 simulated star-field images. Each image was rendered using a telescopic field of view (FOV) of $0.25^\circ$ and simulates real sky-field data derived from Gaia DR3. Every star in each scene includes a corresponding spectral type label, enabling full-spectrum verification. The dataset spans images with between 4 and 20 stars, evenly distributed across buckets \{4, 5, 6, 7, 8, 9, 10, 15, 20\}, allowing us to systematically assess solver robustness under varying detection conditions in Table \ref{tab:fit_rates}.

\subsection{Fit Rate Analysis}

We define a \textit{fit} as a successful astrometric solution, determined by HS-ANET returning a correct World Coordinate System (WCS) alignment. The fit rate at each star count level represents the proportion of images that yielded a correct solution.
As seen in Figure \ref{fig:line_graph}, HS-ANET significantly improves solution robustness, particularly in star-sparse regimes. At 7 detected stars per image, the original pipeline achieves only an 11.3\% fit rate, while HS-ANET reaches 75.3\%. Below this threshold, the original system fails entirely, while HS-ANET continues to provide reliable fits in over 60\% of cases with only 6 stars, and over 40\% with just 5 stars.

\subsection{Normalized AUC Comparison}

To summarize performance across the full star-count range, we compute the \textit{normalized area under the curve (AUC)} using the trapezoidal rule. This metric reflects overall solver robustness and performance efficiency in Table \ref{tab:auc}.

\section{Discussion}

Our results demonstrate that incorporating spectral type information into the astrometric verification process yields substantial improvements in narrow field astrometry with a low number of available stars. HS-ANET achieves higher fit rates than the baseline solver across all star-count bins, and the difference is most pronounced in the 5–10 star range, where the original system struggles or fails entirely. Spectral type offers an additional axis of discriminative power that allows the verification stage to more confidently accept correct hypotheses and reject false ones, even when geometric constraints alone are insufficient.

The improvement in normalized AUC from 0.361 to 0.669 for sparse star regimes provides a compact summary of this gain. We demonstrate better performance at a few thresholds, but a consistent shift toward successful fits across a wider range of input conditions. In operational contexts, particularly with narrow field-of-view telescopes, this means more images can be resolved to accurate sky positions, reducing mission failure rates and increasing measurement reliability.

We also find that the optimal verification parameters $\tau$ and $\lambda$ vary with star density. This suggests future work could benefit from adaptive or learned parameterization, perhaps using meta-learning or a small neural backend trained on scene-level statistics to tune the decision model dynamically.

While our experiments are based on simulated skyfields with Gaia DR3-derived metadata, the methodology is directly transferable to real observations from hyperspectral-capable telescopes. Future work will explore calibration robustness, catalog mismatches, and real-time applicability of this approach.

\section{Conclusion}

We presented HS-ANET, a hyperspectral-enhanced extension to the Astrometry.net pipeline that incorporates stellar spectral type into the astrometric verification process. By augmenting the traditional positional likelihood model with a spectral consistency term, HS-ANET achieves higher or equal fit rates compared to the baseline solver, particularly in scenarios with limited star detections. 

Experiments conducted on 958 synthetic skyfield images derived from Gaia DR3 demonstrate that HS-ANET achieves a normalized AUC of 0.669, compared to 0.361 for the original solver on sparse star regimes. Notably, HS-ANET remains effective with as few as five stars per image, where the baseline fails entirely.

These results highlight the value of hyperspectral information for improving geometric inference in astrometry, especially for narrow-field telescopes. Hyperspectral imagery, combined with algorithms like HS-ANET, has potential to amplify the astrometric precision of ground-based sensors and make strides in Space Traffic Management.

\bibliographystyle{IEEEbib}
\bibliography{strings,refs}

\end{document}